\documentclass[useAMS,usenatbib]{mn2e}
\usepackage{times}
\usepackage{graphicx}
\newcommand{\ds}{\displaystyle}

\usepackage{color}

\title[Chaos in navigation satellite orbits]{Chaos in navigation satellite orbits caused by the perturbed motion of the Moon}
\author[A. J. Rosengren et al.]
{Aaron J. Rosengren,$^{1}$\thanks{E-mail: a.rosengren@ifac.cnr.it} 
Elisa Maria Alessi,$^{1}$
Alessandro Rossi$^{1}$ \and and Giovanni B. Valsecchi$^{1,2}$\\
$^{1}$IFAC-CNR, Via Madonna del Piano 10, 50019 Sesto Fiorentino (FI), Italy\\
$^{2}$IAPS-INAF, Via Fosso del Cavaliere 100, 00133 Roma, Italy}
\begin{document}

\date{}

\pagerange{\pageref{firstpage}--\pageref{lastpage}} \pubyear{2015}

\maketitle

\label{firstpage}

\begin{abstract}
Numerical simulations carried out over the past decade suggest that the orbits of the Global Navigation Satellite Systems are unstable, resulting in an apparent chaotic growth of the eccentricity. Here we show that the irregular and haphazard character of these orbits reflects a similar irregularity in the orbits of many celestial bodies in our Solar System. We find that secular resonances, involving linear combinations of the frequencies of nodal and apsidal precession and the rate of regression of lunar nodes, occur in profusion so that the phase space is threaded by a devious stochastic web. As in all cases in the Solar System, chaos ensues where resonances overlap. These results may be significant for the analysis of disposal strategies for the four constellations in this precarious region of space. 
\end{abstract}

\begin{keywords}
celestial mechanics -- chaos -- methods: analytical -- methods: numerical --  planets and satellites: dynamical evolution and stability --- planets and satellites: general.
\end{keywords}

\section{Introduction}

Space debris---remnants of past missions, satellite explosions, and collisions---is a phenomenon that has existed since the beginning of the space age; however, its significance for space activities, in particular the increasing impact risks posed to space systems, has been realised only in the past few decades \citep{dKbC78,aR99,jLnJ06}. The proliferation of space debris has motivated deeper and more fundamental analysis of the long-term evolution of orbits about Earth \citep{sB01a,sB01b,aCcG14}. Orbital resonances are widespread within this system as a whole \citep{sH80}, but particularly so amongst the medium-Earth orbits (MEOs) of the navigation satellites in the region of semimajor axes between 4 and 5 Earth radii, and a clear picture of their nature is of great importance in assessing debris mitigation measures \citep{eA14}. Indeed, the discovery that the recommended graveyard orbits of these satellites, located several hundred kilometres above the operational constellations, are potentially unstable has led to a new paradigm in post-mission disposal---one that seeks to cleverly exploit these dynamical instabilities and the associated eccentricity growth for re-entry and destruction within the Earth's atmosphere \citep{aJrG02,cC04,aR08,fD11}. Previous studies have already noted the connection between the origin of the long-timescale instabilities in the MEO region and a resonance phenomenon involving Earth oblateness and lunisolar perturbations, yet very little attention has been given to a true physical explanation of the erratic behaviour. To identify the source of orbital instability, we investigated the main resonant structures that organise and govern the long-term orbital motion of navigation satellites. This paper clarifies the fundamental role played by secular resonances and reveals the significance of the perturbed motion of the Moon.  

\section{Analysis}

Orbital resonances are ubiquitous in the Solar System and are harbingers for the onset of dynamical instability and chaos \citep{mD95,mL01,aM02,jC09,kT10,eA14_luna,yLyW14}. Our knowledge of such phenomena in celestial mechanics comes mainly from studies on asteroid dynamics---attempts to describe the stunningly complex dynamical structure of the asteroid belt since the discovery of the Kirkwood gaps \citep{aM02,kT10}. The orbital structure of the main belt is divided by unstable regions, narrow gaps in the distribution of the asteroidal semimajor axes at 2.5 and 2.8 astronomical units (AU) from the Sun, where the orbital periods (or mean motions) of the asteroids are commensurable to that of Jupiter. The outer boundary is marked by the Jovian 2:1 mean-motion resonance at 3.3 AU. Secular resonances, which involve commensurabilities amongst the slow frequencies of orbital precession of an asteroid and the planets, define the inner edge of the main belt near 2.1 AU and demarcate the recognised subpopulations (asteroid families and groups). Secular resonances also occur embedded within the libration zones of the prominent Jovian mean-motion resonances, and it is precisely the coupling of these two main types of resonance phenomena that generates widespread chaos \citep{mL01,aM02,kT10}. The creation of the Kirkwood gaps is thereby associated with a chaotic growth of the asteroids' eccentricities to planet-crossing orbits, which leads to the removal from these resonant regions by collisions or close encounters with the terrestrial planets. The same physical mechanism that sculpted the asteroid belt---the overlapping of resonances---is responsible for the emergence of chaos in the Solar System on all astronomical scales, from the dynamics of the trans-Neptunian Kuiper belt and scattered disk populations of small bodies \citep{mD95} to the apparent perfectly regulated clockwork motion of the planets \citep{yLyW14}.

With the advent of artificial satellites and creation of space debris, similar dynamical situations, every bit as varied and rich as those in the asteroid belt, can occur in closer proximity to our terrestrial abode \citep{pM59,sH80,sB01b}. The dynamical environment occupied by these artificial celestial bodies is subject to motions that are widely separated in frequency: the earthly day, the lunar month, the solar year, and various precession frequencies ranging from a few years to nearly 26,000 years for the equinoxes. A vast and hardly surveyable profusion of perturbations originates from the gravitational action of the Sun on the Earth-Moon system. Primary among these irregularities is the regression the Moon's line of nodes with a period of about 18.61 years, and a progression of the lunar apsidal line with a period of roughly 8.85 years. The provision of frequencies in the Earth-Moon-Sun system gives rise to a diverse range of complex resonant phenomena associated with orbital motions \citep{sH80}. Indeed, past efforts to identify and classify the resonances that generate the orbital instability in the MEO region have been wholly obscured by the abundance of frequencies in the orbital and rotational motions of the Earth and of the Moon, and the great disparity of timescales involved \citep{fD11,tB14,aCcG14}.

While the orbits of most artificial satellites are too low to be affected by mean-motion commensurabilities, even with the Earth's moon \citep{dD14}, there exists a possibility of a somewhat more exotic resonance involving a commensurability of a secular precession frequency with the mean motion of the Sun or the Moon \citep{gC62,mH81,sB01a}. Among the more compelling of these semi-secular commensurabilities is the evection resonance with the Sun, where the satellite's apsidal precession rate produced by Earth's oblateness equals the Sun's apparent mean motion, so that the line of apsides follows the Sun. This is the strongest of the semi-secular apsidal resonances \citep{pM59,sB01b}, and its significance has even been stressed in the dynamical history of the lunar orbit after its hypothesised formation from an impact-generated disk \citep{eA14_luna}, as well as in the dynamics of the Saturnian ring system with the larger, more distant moons, Titan and Iapetus, playing the part of the Sun \citep{jC09}. It is therefore not surprising that some authors \citep{fD11,tB14} have suggested a possible link between these semi-secular resonances and the irregular eccentricity growth in the orbits of the navigation satellites. While it is true that such resonances exist for a wide variety of eccentricities and inclinations, they are only of trivial importance in these regions, as they mainly occur at much lower semimajor axes \citep{sB01b}. 

Satellite orbits resonant with respect to the perturbing influence of the Earth's gravitational field are of particular importance for the navigation satellite systems \citep{tEkH97,tE02,aR08,aCcG14}, and such resonance problems have stood in the foremost rank of astrodynamical research work \citep{sB01b}. Undoubtedly, the most celebrated resonance phenomenon in artificial satellite theory is the critical inclination problem, which involves a commensurability between the two degrees of freedom (apsidal and nodal motions) of the secular system \citep{mH81,fDaM93,sTtY14}. The extreme proximity to these critical inclinations, especially in the case of the European Galileo and Russian GLONASS constellations \citep{aR08,eA14}, induced researchers, very early, to associate the origin of the instabilities observed in numerical surveys to these resonances \citep{aJrG02,cC04}. However, these inclination-dependent-only resonances are generally isolated \citep{gC62,sH80} and thereby exhibit a well-behaved, pendulum-like motion, which, when considered alone, cannot explain the noted chaotic behaviours. The second class of geopotential resonances arises from the slight longitude-dependence of the geopotential, where the satellite's mean motion is commensurable with the rotation of the Earth. Under such conditions, the longitudinal forces due to the tesseral harmonics continually perturb the orbit in the same sense and produce long-term changes in the semimajor axis and the mean motion. For the inclined, slightly eccentric orbits of the navigation satellites, where these commensurabilities abound, the tesseral harmonics exhibit a multiplet structure, akin to mean-motion resonances, whereby the interaction of resonant harmonics can overlap and produce chaotic motions of the semimajor axis \citep{tEkH97,tE02}. These effects, however, are localised to a narrow range of semimajor axis (tens of kilometres) and are of much shorter period than secular precession \citep{tEkH97,aCcG14}; consequently, tesseral resonances will not significantly affect the long-term orbit evolution over timespans of interest.    

\section{Results and Discussion}

These considerations have led us to investigate the role of lunar secular resonances in producing chaos and instability among the navigation satellites. We treat only the secular resonances of the lunar origin, since, despite recent implications to the contrary \citep{tB14}, the solar counterparts will require much longer timespans than what interests us here for their destabilising effects to manifest themselves \citep{tE02}---a consequence of the disparity of several orders of magnitude between the orbit precession periods of the Earth and the Moon. As a basis for our calculations we have used a convenient trigonometric series development by \citet{sH80}, corresponding largely to a harmonic analysis of the perturbations. The lunar disturbing function is developed as a Fourier series of complicated structure, whose arguments are combinations of the orbital phase and orientation angles of the satellite and the Moon, and whose coefficients depend on the size and shape (semimajor axes and eccentricities) of their orbits and the inclinations. Two considerable simplifications are possible, reducing this rather formidable expression in a marked degree. For satellites whose semimajor axis does not exceed one tenth of the Moon's distance from the Earth, we can truncate the series to second order in the ratio of semimajor axes, so that the lunar potential is approximated with sufficient accuracy by a quadrupole field \citep{pM61,fDaM93,aRdS13,sTtY14}. To study the secular interactions, the short periodic terms of the disturbing function, depending on the mean anomalies of both the satellite and the Moon, can be averaged out \citep{aM02}. For lack of any mean-motion or semi-secular resonances, such averaging effectively involves discarding all terms which depend on these fast orbital phases. Accordingly, the lunar disturbing function reduces to the form:  
\begin{equation}
\begin{array}{l}
	\ds \mathcal{R} 
	= \frac{\mu_M a^2}{a_M^3 (1 - e_M^2)^{3/2}} \sum_{m = 0}^2 \sum_{s = 0}^2
		(-1)^m K_m \frac{(2 - s)!}{(2 + m)!}  \\[1.25em]
	\ds \hspace{21pt}\sum_{p = 0}^2 F_{2,m,p} (i) F_{2,s,1} (i_M) G_{2,p,2p-2} (e) \\[1.25em]
	\ds \hspace{21pt} 
		\left[ (-1)^{2-s} U_2^{m,-s} (\epsilon) \cos \Phi^{+}_{p, m, s} 
		+ U_2^{m,s} (\epsilon) \cos \Phi^{-}_{p, m, s} \right],
\end{array}
\end{equation}
with harmonic angles
\begin{equation}
	\Phi^{\pm}_{p, m, s} = (2 - 2 p) \omega + m \Omega \pm s \Omega_M \pm \frac{\pi}{2} (s + m), 
\end{equation}
where $m$, $s$, and $p$ are integers and the quantity $K_m$ is such that $K_0 = 1$ and $K_m = 2$ for $m > 0$. The semimajor axis $a$, eccentricity $e$, inclination $i$, longitude of ascending node $\Omega$, and argument of pericentre $\omega$ are the satellite's orbital elements relative to the celestial equator. The ecliptic orbital elements of the Moon are denoted by subscript $M$, its gravitational mass by $\mu_M$, and $\epsilon$ is the Earth's obliquity. The functions $F$ and $G$ are the modified Allan inclination function and Hansen coefficient, respectively; and the function $U$ accounts for the fact that the Moon's orbit is referred to the ecliptic \citep[see][for more details]{sH80}:
\begin{equation}
\begin{array}{l}
	\ds U_2^{m, \pm s} 
	= \frac{(-1)^{m \mp s}}{(2 \pm s)!} (\cos \epsilon/2)^{m \pm s} 
		(\sin \epsilon/2)^{\pm s - m} \\[1.25em]
	\ds \hspace{39pt} \frac{\mathrm{d}^{2 \pm s}}{\mathrm{d}\, Z^{n \pm s}} 
		\left\{ Z^{2 - m} (Z - 1)^{n + m} \right\},
\end{array}
\end{equation}
where $Z = \cos^2 \epsilon/2$. Such a choice of reference planes permits us to consider $i_M$ as roughly constant and the variation in $\Omega_M$, largely caused by the disturbing action of the Sun, as approximately linear with period 18.61 years \citep{sH80}. The rather remarkable fact that the Moon's perigee does not explicitly appear in the quadrupolar expansion of the secular problem has been known for some time \citep{pM61}, and belies the recent assertion \citep{tB14} that the position of the lunar perigee gives rise to resonance effects on navigation satellite orbits. 

\begin{figure}
\centering
\includegraphics[scale=0.65]{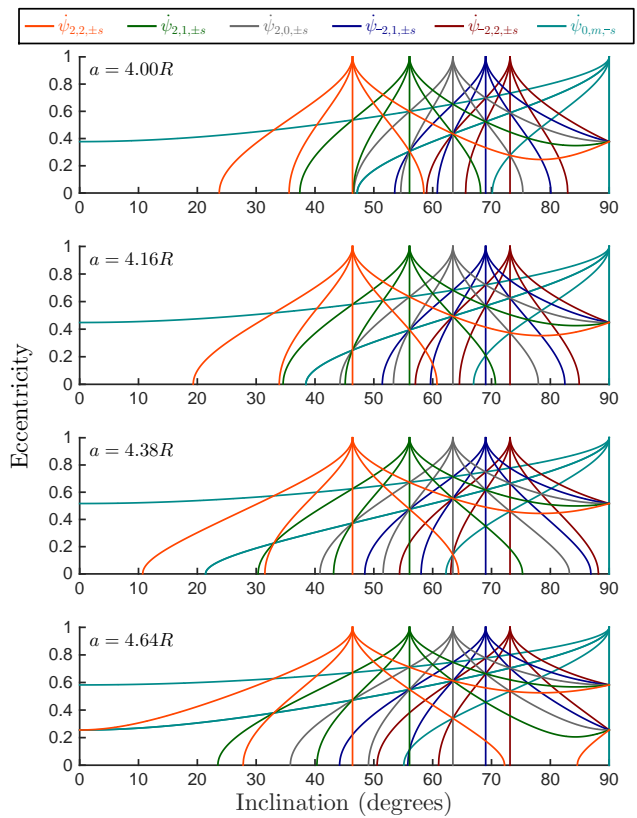}
\caption{The resonant structure of the medium-Earth orbits of the four navigation constellations: GLONASS ($a=4.00R$), GPS ($a=4.16R$), BeiDou ($a=4.38R$), and Galileo ($a=4.64R$). Each curve represents the location of the exact resonance of the form $\dot\psi_{2 - 2 p, m, \pm s} = (2 - 2 p) \dot\omega + m \dot\Omega \pm s \dot\Omega_M = 0$, where the effects of perturbations on $\omega$ and $\Omega$ other than the $J_2$ harmonic have been neglected. For each group of curves, the vertical line corresponds to the inclination-dependent-only lunisolar secular resonances ($s = 0$), while the other curves are obtained for $s = -2, -1, 1, 2$. \citet{tEkH97} were the first to identify this devious network of resonances for the GPS constellation; yet the exact nature of its consequences has not been hitherto explored. These resonances generally form the skeletal structure of the phase space, tracing the topological organisation of the manifolds on which the chaotic motions take place \citep{aM02}. Chaos is produced by the overlapping of the closely-spaced resonant harmonics \citep{bC79}.}
\label{figure1}
\end{figure}

A lunar secular resonance occurs when
\begin{equation}
	\label{eq:res_con}
	\dot\psi_{2 - 2 p, m, \pm s} = (2 - 2 p) \dot\omega + m \dot\Omega \pm s \dot\Omega_M \approx 0; 
\end{equation}
that is, when a specific linear combination of the secular precession frequencies vanishes. For the Moon, the rate of change of $\Omega_M$ is approximately $-0.053$ deg/d. The oblateness precession completely overshadows the lunisolar effects and other perturbing influences so that $\omega$ and $\Omega$ have an essentially linear time dependence,\footnote{This approximation is satisfactory when compared with the rigorous theory. Neglect of the lunisolar perturbations on the the frequencies of nodal and apsidal precession sets an upper limit to the radius of the orbit for which the theory is valid; on the other hand, the period of precession must be appreciably longer than a year for the double averaging procedure to be justifiable, which sets a lower limit to the orbital radius. For these reasons, the analysis is most useful in the region of semimajor axes between 4 and 5 Earth radii.} with secular rates given by the classical expressions \citep{gC62}:
\begin{equation}
\label{eq:oblateness}
\begin{array}{l}
	\ds \dot\omega = \frac{3}{4} J_2 n \left( \frac{R}{a} \right)^2 \frac{5 \cos^2 i - 1}{(1 - e^2)^2}, \\[1.25em]
	\ds \dot\Omega = -\frac{3}{2} J_2 n \left( \frac{R}{a} \right)^2 \frac{\cos i}{(1 - e^2)^2},
\end{array}
\end{equation}
where $J_2$ is the second zonal harmonic coefficient of the geopotential, $R$ is the mean equatorial radius of the Earth, and $n$ is the satellite's mean motion. Equations~(\ref{eq:res_con}) and (\ref{eq:oblateness}) define surfaces of secular resonances in the space of the orbital elements $a$, $e$, and $i$. As the semimajor axis is secularly invariable after averaging, the commensurability condition defines curves of secular resonances in eccentricity and inclination phase space. The interaction of various lunar harmonics produces an exceedingly complicated network of resonances---courtesy of the Earth's oblateness---becoming particularly dense near the inclinations of the navigation satellite orbits (Fig.~\ref{figure1}). Each of the critical inclinations of the geopotential resonances (corresponding to $s = 0$\footnote{\citet{sH80} classifies these commensurabilities as inclination-dependent-only lunisolar secular resonances, under the assumption that the $J_2$ harmonic in the geopotential produces the dominant change in the satellite's perigee and node \citep[see also][]{gC62,sB01b}.}) split into a multiplet of resonant curves, emanating from unity eccentricity.\footnote{To depict such curves up to $e = 1$ is only theoretical, as the satellites will re-enter Earth's atmosphere when $e > 1 - R/a$. Nevertheless, we are mainly interested here in the early stages in the development of chaos.} Each of the resonance curves in Fig.~\ref{figure1} has its own pendulum-like structure, with a characteristic domain of libration. When the separation between nearby resonances becomes similar to their libration widths, the resonant critical angles of the trajectories in these overlapping regions may switch irregularly between libration and circulation; this alternation is a hallmark of chaos \citep{aM02,yLyW14}.

\begin{figure}
\centering
\includegraphics[scale=0.65]{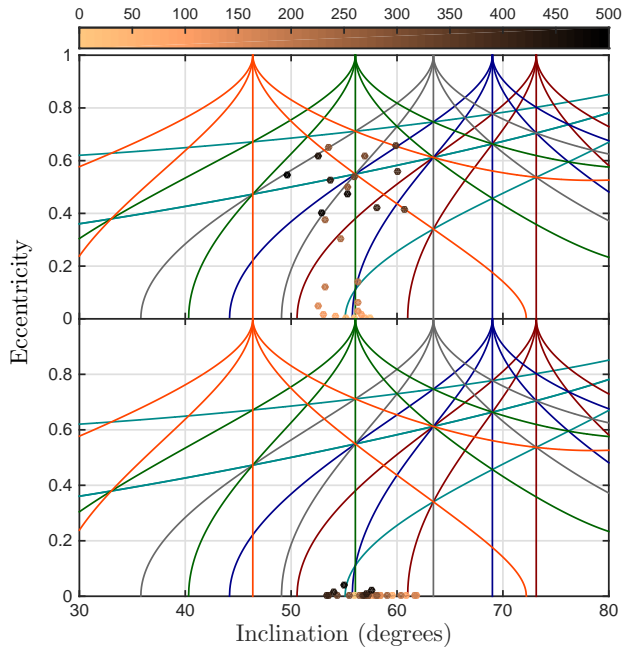}
\caption{Dynamical evolution of two Galileo-like orbits superimposed on the background of complex resonant curves. The initial epoch of the simulations is 26 February 1998, which determines the initial dynamical configuration of the Earth-Moon-Sun system, and the orbits were propagated for 500 years (copper-tinge sequence of discrete points) using an accurate, non-singular, singly-averaged model of oblateness and lunisolar perturbations, develop previously by one of the authors \citep{aRdS13}. Both representative Galileo orbits initially have $a=29,600$ km, $e=0.001$, $i=56^\circ$, $\omega = 30^\circ$, but their orbital planes are separated by $120^\circ$ with $\Omega=240^\circ$ (top) and $\Omega=120^\circ$ (bottom), respectively. We consider the discrete dynamics of the system by taking a subsequent snapshot of the motion every lunar nodal period (a stroboscopic map), which gives a faithful picture of the general character of the trajectories. The motions with initial conditions in the overlapping regions can be both chaotic and regular, depending on the initial orientation angles.}
\label{figure2}
\end{figure}

\begin{figure}
\centering
\includegraphics[scale=0.65]{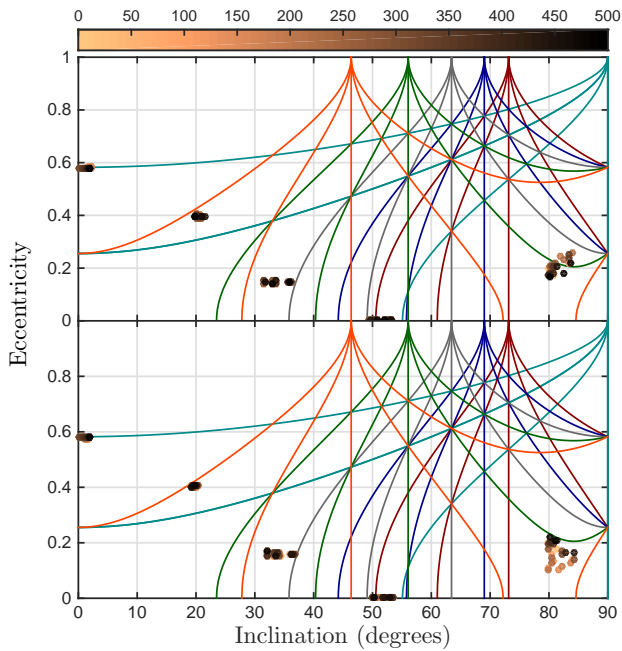}
\caption{Dynamical evolution of five orbits with initial inclinations and eccentricities in non-overlapping resonance regions superimposed on the background of secular resonant curves. The initial epoch of the simulations is 26 February 1998, and the orbits were propagated for 500 years (copper-tinge sequence of discrete points) using an accurate, non-singular, singly-averaged model of oblateness and lunisolar perturbations \citep{aRdS13}. All orbits initially have $a=29,600$ km, $\omega = 30^\circ$, but their orbital planes are separated by $120^\circ$ with $\Omega=240^\circ$ (top) and $\Omega=120^\circ$ (bottom), respectively. We consider the discrete dynamics of the system by taking a subsequent snapshot of the motion every lunar nodal period (a stroboscopic map). The motions with initial conditions in the non-overlapping regions are generally regular; chaotic motion can only exist close to the separatrices of the isolated resonances \citep{aM02}.}
\label{figure3}
\end{figure}

To gain an understanding of the physical significances of these resonances, it it important to develop analytical models that accurately reflect the true nature of the resonant interactions. The simplest physical model for the study of the secular and resonant motion in MEO consists of the quadrupole order of approximation for the Earth, lunar, and solar gravitational potentials, but with the Moon's motion represented by a precessing and rotating elliptical orbit. The singly-averaged model forms a nonautonomous Hamiltonian two degrees-of-freedom system, depending periodically on time through the MoonÕs perturbed motion \citep{aRdS13}. Numerical simulations of this model show that the regions where two or more neighbouring lunar secular resonances interact exhibit chaos and large-scale excursions in eccentricity (Fig.~\ref{figure2}), in agreement with theoretical predictions based on the resonance overlap criterion \citep{bC79}. Chaotic diffusion is mediated by the web-like structure of secular resonances, which permeate the phase space and allows an initially circular orbit to become highly eccentric, as revealed in a Poincar\'{e} map of the nonautonomous, periodically-perturbed two degrees-of-freedom system, obtained by sampling the flow stroboscopically once per lunar nodal period. Once a dynamical instability sets in, the subsequent evolution is highly chaotic and unpredictable in detail. The resonances are the preferential routes for chaotic diffusion, as the trajectory jumps from one resonance domain to the other. In any case, the perigee will drive toward the ground, and the satellite is doomed to destruction. The chaotic zones defined by the regions of overlapping resonances do not preclude the existence of regular trajectories embedded within it. Indeed, the character of the motion depends sensitively upon the initial orientation angles of the satellite and the initial lunar node (Fig.~\ref{figure2}). The second orbit (bottom) suffers only small-amplitude variations in eccentricity and inclination, and is apparently regular, at least on the timescale of 500 years. Figure~\ref{figure3} shows that the absence of resonance overlapping generally guarantees the local confinement of the motion. A large amount of simulations, in which the initial perigee and nodes were varied, were performed and further confirm this result: satellites with initial conditions in the non-overlapping region exhibit regular, bounded motions. For the full system one might expect that a coupling of the tesseral and lunar resonance phenomena would produce a slow diffusion in phase space, by which orbits can explore large regions jumping from one resonance to another. A discussion of this aspect of chaotic diffusion has been given by others \citep{tEkH97,tE02}, and only occurs over millennia timescales.

It should be reiterated that the resonant curves presented in Fig.~\ref{figure1} show the regions in the inclination-eccentricity phase space for which chaotic orbits can be found and explain why chaotic orbits manifest in these regions only. It gives, however, no information about which initial angles ($\omega$, $\Omega$, and $\Omega_M$) will lead to chaos. The structure of the whole phase space is far from being fully understood and clearly warrants greater theoretical development. Future work will present more complicated maps, which allow for the global visualisation of the geometrical organisation and coexistence of chaotic and regular motion.

\section{Conclusion}

In the inner Solar System, overlapping secular resonances have been identified as the origin of chaos in the orbits of the terrestrial planets \citep{mL01,aM02,yLyW14}. We have demonstrated that the same underlying dynamical mechanism responsible for the eventual destabilisation of Mercury, and recently proposed to explain the orbital architecture of extra-solar planetary systems \citep{yLyW14}, is at the heart of the orbital instabilities of seemingly more mundane celestial bodies---the Earth's navigation satellites. The occurrence and nature of the secular resonances driving these dynamics depend chiefly on one aspect of the Moon's perturbed motion, namely, the regression of the line of nodes. The decisive significance of this fact on space debris mitigation will be emphasised in a later paper and its relation to the design of disposal strategies in MEO \citep{eA14} will be formulated there more precisely. The precarious state of these constellations, perched on the threshold of instability, makes it understandable why all efforts to define stable graveyard orbits, especially in the case of Galileo, were bound to fail; the region is far too complex to allow of an adoption of the simple geosynchronous disposal strategy. A full understanding of the nature and consequences of the chaos in these environments would have certainly helped in the early design phases of the constellations.

Active debris removal, apart from the daunting obstacles in engineering that this represents, is currently seen \citep{jLnJ06} as the only viable option to prevent the self-generating Kessler syndrome phenomenon \citep[collisional cascading; see][]{dKbC78} from occurring in low-Earth orbit, the most densely populated orbital environment \citep{aR99}. But as we are still remarkably ignorant of the locations and consequences of most resonances in near-Earth space, such drastic measures may require a reassessment. This concerns particularly the question as to whether strong instabilities exist, whose destabilising effects occur on decadal timescales, that can be exploited to effectively clear these regions of space from any future collision hazard. Indeed, the process of dynamical clearing of resonant orbits is well illustrated by the paucity of asteroids observed in the Kirkwood gaps \citep{kT10}. 

\section*{Acknowledgments}

The present form of the manuscript owes a great deal to the reviewer, Alessandra Celletti, of the University of Roma Tor Vergata; by her various valuable suggestions, many concepts and results have been clarified. We also owe special thanks to Florent Deleflie and J\'{e}r\^{o}me Daquin, of the IMCCE/Observatoire de Paris, and Kleomenis Tsiganis, of the Aristotle University of Thessaloniki, for many insightful and motivating conversations. This work is partially funded by the European CommissionÕs Framework Programme 7, through the Stardust Marie Curie Initial Training Network, FP7-PEOPLE-2012-ITN, Grant Agreement 317185. Part of this work was performed in the framework of the
ESA Contract No. 4000107201/12/F/MOS ``Disposal Strategies Analysis for MEO Orbits''.

\bibliography{mnras_refs}

\bibliographystyle{mn2e}

\bsp

\label{lastpage}

\end{document}